\documentclass{PoS}

\newcommand{\Fi}[1]{Fig.~\ref{#1}}

\newcommand{\agev}{\mbox{~$A$GeV}}
\newcommand{\gevc}{\mbox{GeV$/c$}}

\newcommand{\rb}[1]{\mbox{\textrm{\scriptsize #1}}}
\newcommand{\rbt}[1]{\mbox{\textrm{\tiny #1}}}

\newcommand{\sqrts}{\ensuremath{\sqrt{s_{_{\rbt{NN}}}}}}

\newcommand{\lam}{\ensuremath{\Lambda}}
\newcommand{\lab}{\ensuremath{\bar{\Lambda}}}
\newcommand{\myphi}{\ensuremath{\phi}}
\newcommand{\pimin}{\ensuremath{\pi^-}}

\newcommand{\kmin}{\ensuremath{\textrm{K}^-}}
\newcommand{\kplus}{\ensuremath{\textrm{K}^+}}

\newcommand{\kzero}{\ensuremath{\textrm{K}^{0}}}
\newcommand{\kzerob}{\ensuremath{\bar{\textrm{K}}^{0}}}
\newcommand{\sigzero}{\ensuremath{\Sigma^{0}}}
\newcommand{\sigzerob}{\ensuremath{\bar{\Sigma}^{0}}}
\newcommand{\sigpm}{\ensuremath{\Sigma^{\pm}}}
\newcommand{\sigpmb}{\ensuremath{\bar{\Sigma}^{\pm}}}
\newcommand{\xis}{\ensuremath{\Xi^{0,-}}}
\newcommand{\xisb}{\ensuremath{\bar{\Xi}^{0,+}}}
\newcommand{\ommin}{\ensuremath{\Omega^-}}

\newcommand{\omplus}{\ensuremath{\bar{\Omega}^+}}

\newcommand{\mypt}{\ensuremath{p_{\rb{t}}}}

\newcommand{\mt}{\ensuremath{m_{\rb{t}}}}

\newcommand{\meanmtm}{\ensuremath{\langle m_{\rb{t}} \rangle - m_{\rb{0}}}}

\newcommand{\nwound}{\ensuremath{\langle N_{\rb{w}} \rangle}}

\newcommand{\gams}{\ensuremath{\gamma_{\rb{S}}}}

\newcommand{\kppip}{\ensuremath{\langle \textrm{K}^+ \rangle / \langle \pi^+ \rangle}}
\newcommand{\kmpim}{\ensuremath{\langle \textrm{K}^- \rangle / \langle \pi^- \rangle}}
\newcommand{\lampi}{\ensuremath{\langle \Lambda \rangle / \langle \pi \rangle}}
\newcommand{\xipi}{\ensuremath{\langle \Xi^{-} \rangle / \langle \pi \rangle}}
\newcommand{\ompi}{\ensuremath{\langle \Omega^{-} + \bar{\Omega}^{+} \rangle / \langle \pi \rangle}}

\newcommand{\vtwo}{\ensuremath{v_{\rb{2}}}}

\title{Review of Structures in the Energy Dependence of Hadronic Observables}

\ShortTitle{Structures in Energy Dependencies}

\author{\speaker{C. Blume}\\
        Institut f\"{u}r Kernphysik, J.W.~Goethe Universit\"{a}t, Frankfurt am Main, Germany\\
        E-mail: \email{blume@ikf.uni-frankfurt.de}}

\abstract{The energy dependence of various hadronic observables is reviewed.
The study of their evolution from AGS over SPS to the highest RHIC 
energy reveals interesting features, which might locate a possible
onset of deconfinement. These observables include transverse spectra of 
different particle types and their total multiplicities, as well as
elliptic flow. 
In this context especially the observation of a maximum of the strangeness 
to pion ratio is of particular interest, since on one hand it has been 
predicted as a signal for the onset of deconfinement but on the other hand 
also statistical model calculations exhibit qualitatively similar structures. 
The sharpness of these features is however not reproduced by hadronic scenarios. 
The significance of these structures will be discussed in this contribution. 
Other observables, such as radius parameters from Bose-Einstein correlations, 
on the other hand do not exhibit any structure in their energy dependence.}

\FullConference{The 3rd edition of the International Workshop ---
                The Critical Point and Onset of Deconfinement --- \\
	        July 3-7 2006 \\
		Galileo Galilei Institute, Florence, Italy}

\begin{document}

\section{Introduction}

\begin{figure}[t]
\begin{center}
\begin{minipage}[b]{70mm}
\begin{center}
\includegraphics[height=120mm]{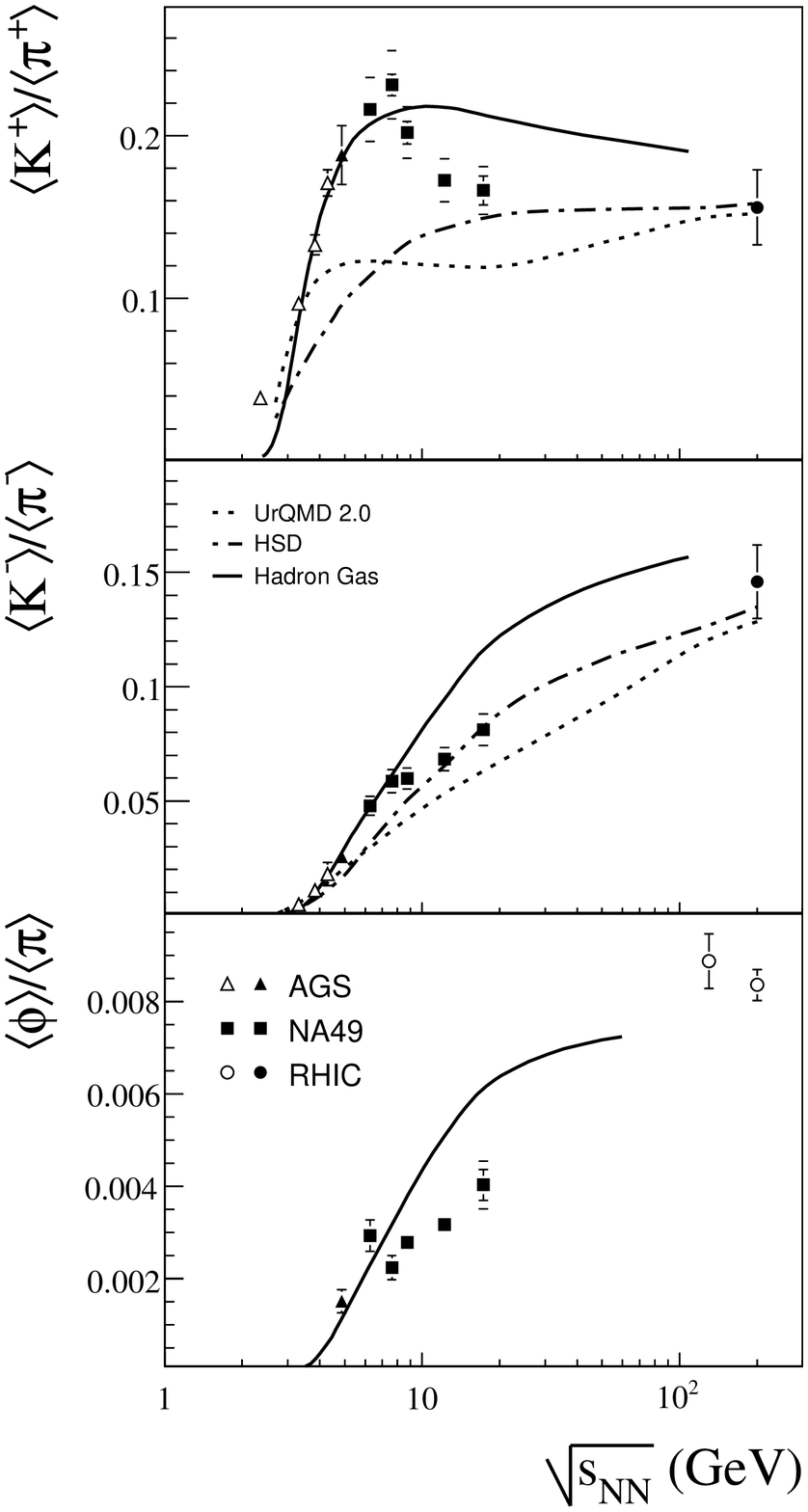}
\end{center}
\end{minipage}
\begin{minipage}[b]{70mm}
\begin{center}
\includegraphics[height=120mm]{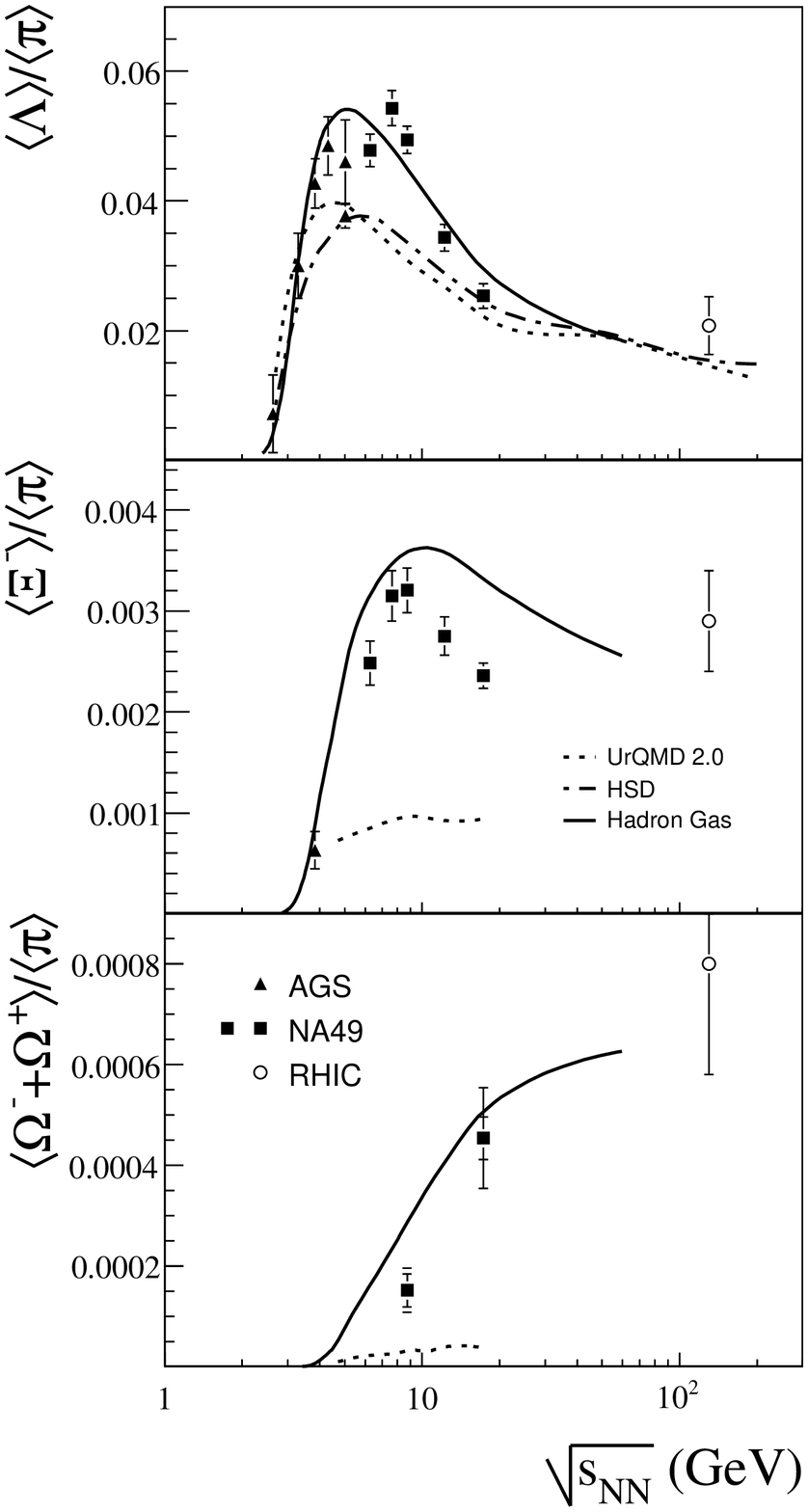}
\end{center}
\end{minipage}
\end{center}
\caption{The energy dependence of the 4$\pi$-yields of
strange hadrons, normalized to the pion yields, in central
Pb+Pb/Au+Au collisions \cite{michi}. The data are
compared to string hadronic models (UrQMD 2.0 \cite{urqmd}: 
dotted lines, HSD \cite{hsd}: dashed-dotted lines) 
and a statistical hadron gas model \cite{pbm}, 
assuming full chemical equilibrium (solid line).}
\label{Fig:Ratios}
\end{figure}

Since the chemical freeze-out points approach the phase boundary line 
as predicted by lattice QCD already at top SPS energies, it is a 
reasonable conjecture that the transition to a quark-gluon plasma is
likely to happen in the SPS energy regime (i.e. $\sqrts = 6 - 17.3$~GeV).
Also the energy density, which is achieved at top SPS beam energies, 
should be sufficiently high for a QGP formation.
Therefore, a study of the energy dependence of heavy ion collisions might
help to identify a possible phase transition and
to localize the center-of-mass energy, where the phase boundary is 
reached first. In such a study one should look for any kind of
``non-smooth'' behaviour in the energy dependence of different observables.
The most prominent examples of such kind of structures have been reported
by the NA49 collaboration \cite{marekqm}: A pronounced maximum in the 
\kppip-ratio, and sudden changes in the energy dependence of the slope
parameters of kaons and of pion production.
In \cite{smes} two of these observations have been predicted as a
signature for the onset of deconfinement.
The questions that will be addressed in this contribution are: Are there any
other structures in the energy dependence of hadronic observables and what
is the significance of any of these?
Hadronic models (transport models like UrQMD \cite{urqmd} and HSD \cite{hsd} 
or statistical models) 
provide an important baseline for a comparsion and help to identify
``trivial'' structures. One important aspect in this context is the
fact that the fireball produced in heavy ion collisions changes its nature 
from being baryon dominated at lower energies to
meson dominated at higher energies. In the framework of a statistical
hadron gas model this transition was located around $\sqrts \approx 8$~GeV
which coincides with a maximum of the relative strangeness production 
\cite{cleymans}. Therefore any kind of structure observed in the data has to
be weighted against structures expected from this hadronic scenario.

\section{Particle Yields}

\begin{figure}[t]
\begin{center}
\begin{minipage}[b]{70mm}
\begin{center}
\includegraphics[height=120mm]{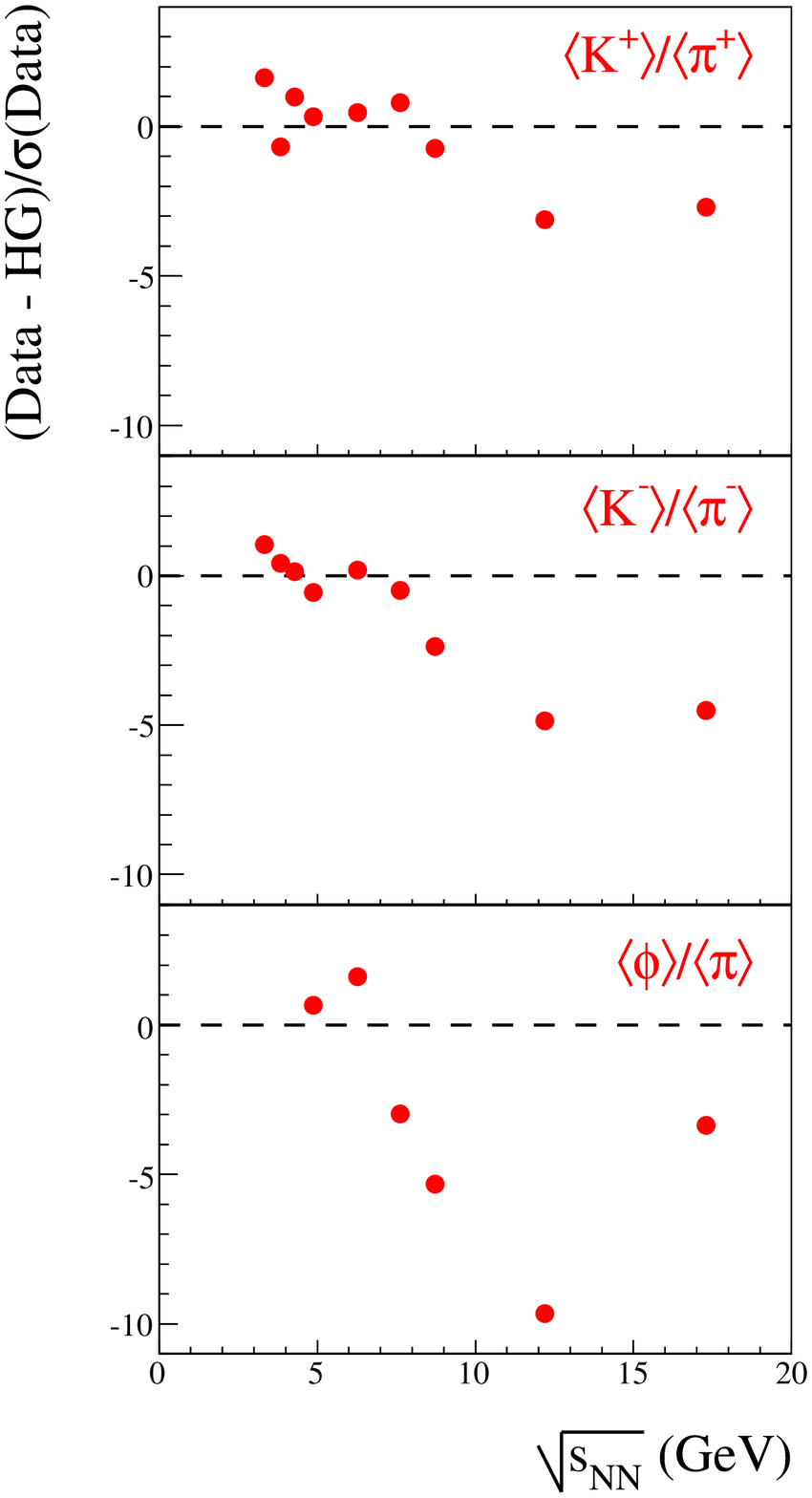}
\end{center}
\end{minipage}
\begin{minipage}[b]{70mm}
\begin{center}
\includegraphics[height=120mm]{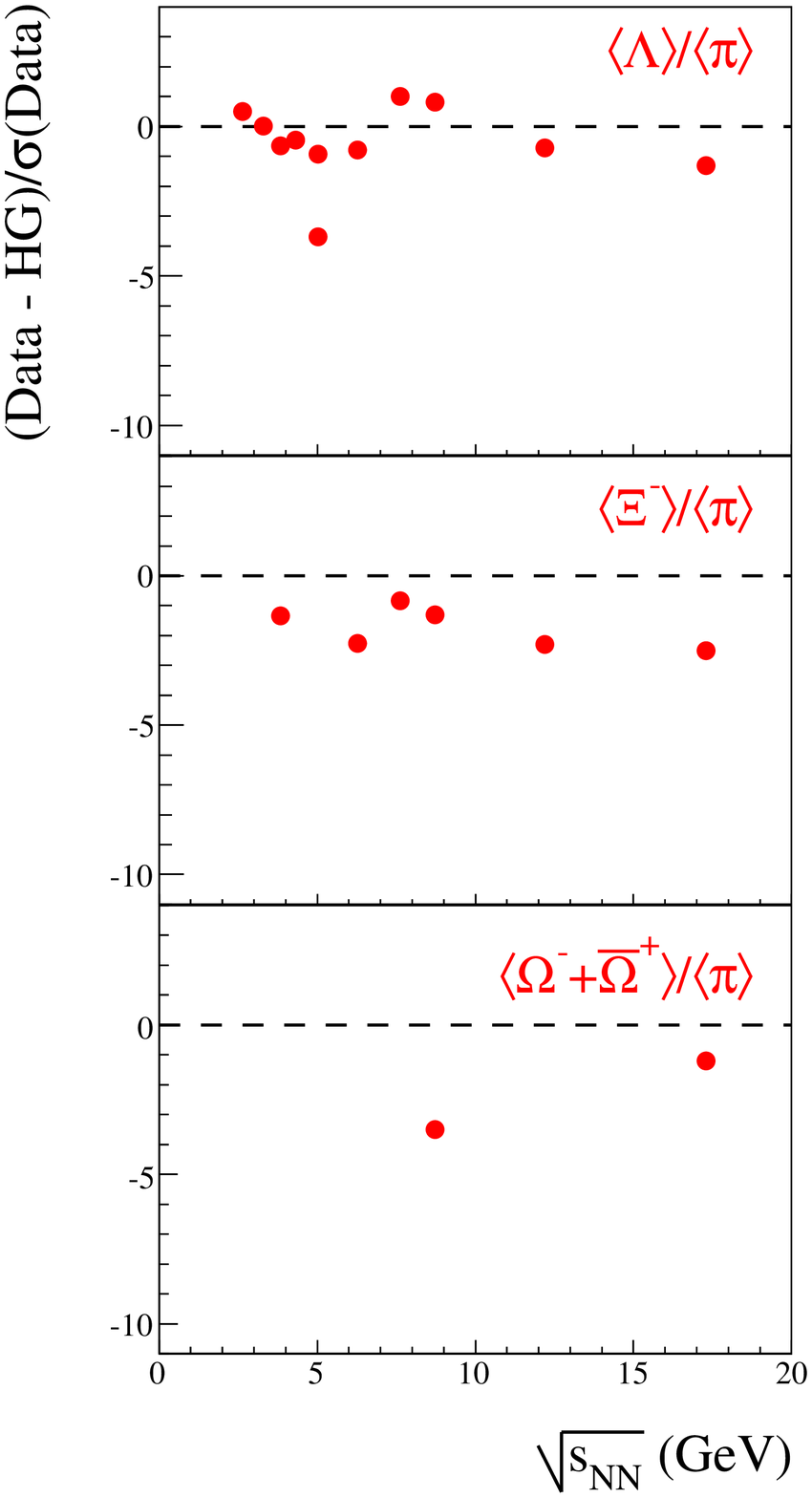}
\end{center}
\end{minipage}
\end{center}
\caption{The difference of the measured particle ratios to the predictions 
of a fully equilibrated hadron gas \cite{pbm} relative to the total error.}
\label{Fig:RatDev}
\end{figure}

\begin{figure}[th]
\begin{center}
\includegraphics[width=110mm]{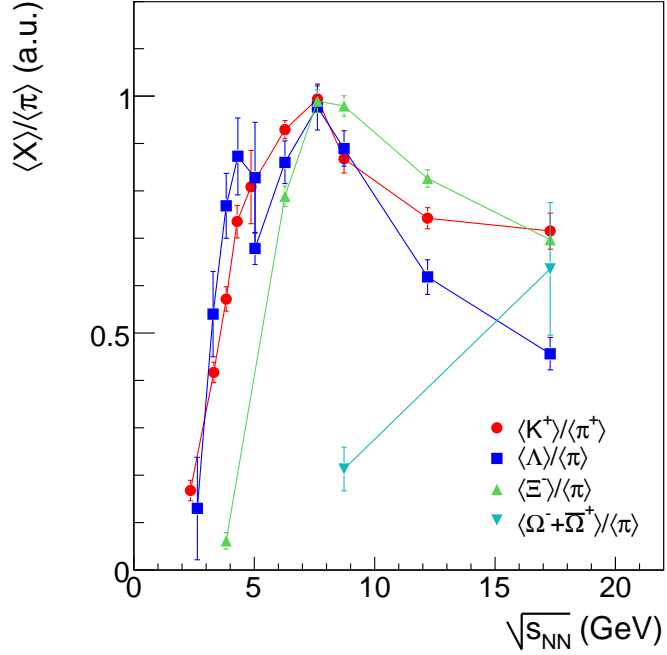}
\end{center}
\caption{The energy dependence of the 4$\pi$-yields of
strange hadrons, normalized to the pion yields, in central
Pb+Pb/Au+Au collisions. The ratios were scaled such that the
maximum value is unity.}
\label{Fig:Horn}
\end{figure}

In \Fi{Fig:Ratios} the energy dependence of the total multiplicities 
for a variety of strange hadrons, normalized to the
pion yield, as measured by NA49 and various AGS and RHIC experiments \cite{michi},
is summarized and compared to model predictions. 
Generally, it can be stated that the string hadronic models UrQMD \cite{urqmd} 
and HSD \cite{hsd} do not provide a good description of the data points.
Especially the $\Xi$ and $\Omega$ production is substantially underestimated
and the maximum in the \kppip\ ratio is not reproduced. 
The basic trend of the \kmpim\ and the \lampi\ ratios is on the other hand
better described by the transport models, although discrepancies remain.  
A statistical hadron gas model assuming full chemical equilibrium \cite{pbm}, 
on the other hand, provides a better overall description of the measurements. 
This kind of model implictly takes into account the already mentioned
transition from a baryon to a meson dominated system. 
However, also here clear differences to the NA49 measurements exist.
These require the introduction of an energy dependent strangeness 
saturation factor \gams\ \cite{becatt}, in order to capture the 
structures in the energy dependence of most particle species (not shown here).

Since the full chemical equilbrium assumption in principle provides a 
relatively well defined theoretical baseline that is also qualitatively
describing the trend of the measurements, it is worthwhile to look at 
the differences in more detail in order to see whether there are any
common features.
These are summarized in \Fi{Fig:RatDev}. As the left panel shows, both, 
charged kaons as well as $\phi$ mesons deviate by more than three times the
total error from the statistical model prediction for $\sqrts > 7 - 8$~GeV.
The situation is less clear for $\Lambda$ and hyperons (\Fi{Fig:RatDev}, 
right panel). While the \lampi\ ratios generally agree within the total error, 
there might be also an indication for a departure
from the full equilibrium assumption for the \xipi\ ratios, although not
as obvious as for the kaons and $\phi$. The \ompi\ ratio, on the other
hand, seems to deviate stronger at the lower beam energy. However, for
a more systematic evaluation of its energy dependence more data at
different energies would be needed. Especially the question, whether 
the difference of the measured \ompi\ ratio to the statistical model 
prediction increases towards even lower energies, or remains close to
it (as it seems to be the case for the \xipi\ ratio) would be of 
particular interest. The first scenario would be expected
if a dynamical equilibration of a heavy particle like the $\Omega$ is
only possible at higher energies where the system is close to the 
phase boundary \cite{pbmjs}. The second would rather indicate that
particle production via the strong interaction always follows the
maximum entropy principle. 

As has been pointed out in \cite{cleymans}, one of the predictions of
the statistical model is that the position of the maxima in 
the energy dependence of the strange particles to pion ratios should
depend on the strangeness content and the baryon number. In \cite{cleymans} 
the following positions are predicted: $\sqrts = 10.8$~GeV ($\textrm{K}^{+} / \pi^{+}$),
5.1~GeV ($\Lambda / \pi$), 10.2~GeV ($\Xi^{-} / \pi^{-}$),
27.0~GeV ($\Omega^{-} / \pi^{-}$).
Figure~\ref{Fig:Horn} summarizes the current situation on the 4$\pi$ ratios, as
also shown in \Fi{Fig:Ratios}.
They were normalized to unity at the maximum in order to allow for a direct
comparison of their positions. While the position of the maximum for the
\kppip\ and \lampi\ ratios seems to coincide, it appears to be shifted a bit
to higher values for the \xipi\ ratio. However, both, the maximum for \kppip\
and for \xipi are at smaller values of \sqrts\ than predicted by the statistical
model. For the \ompi\ ratio no maximum can be observed, but rather a continous
rise (see also right panel of \Fi{Fig:Ratios}). 

\begin{figure}[t]
\begin{center}
\begin{minipage}[b]{70mm}
\begin{center}
\includegraphics[height=70mm]{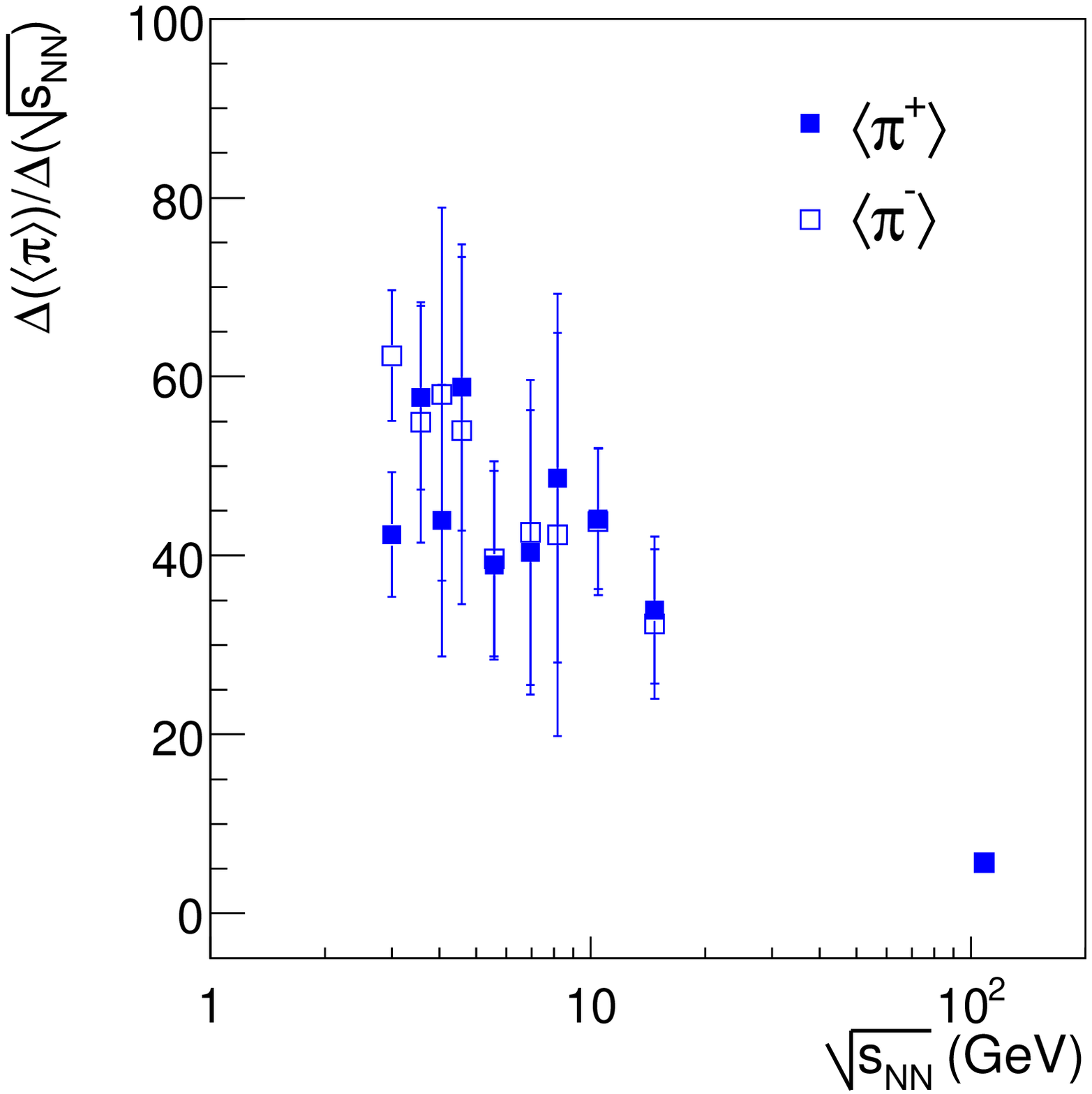}
\end{center}
\end{minipage}
\begin{minipage}[b]{70mm}
\begin{center}
\includegraphics[height=70mm]{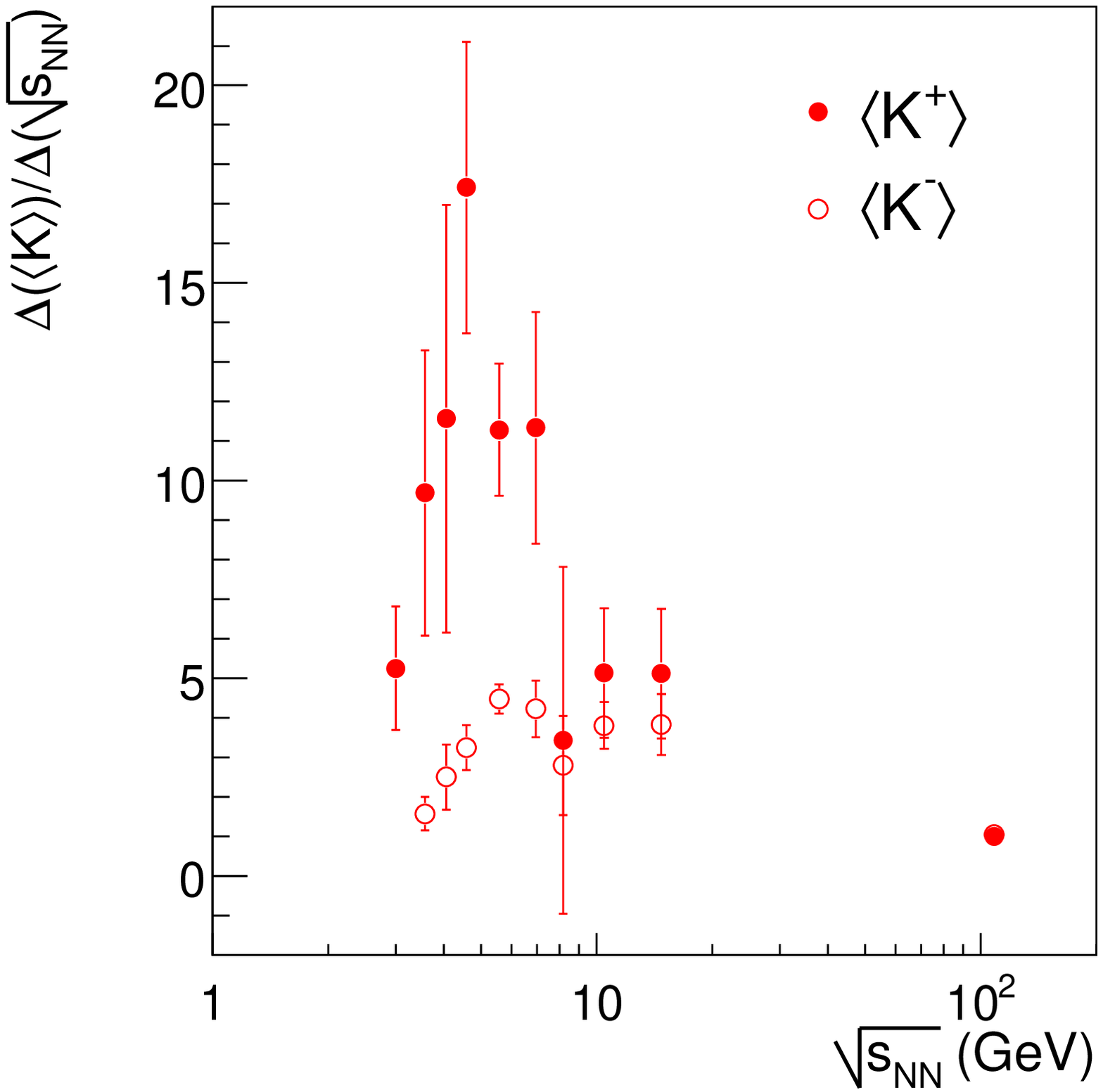}
\end{center}
\end{minipage}
\end{center}
\caption{The local slopes of the energy dependencies of
charged pions (left panel) $\Delta \langle \pi \rangle / \Delta \sqrts$
and charged kaons (right panel) $\Delta \langle \textrm{K} \rangle / \Delta \sqrts$
as a function of \sqrts.}
\label{Fig:KPiDer}
\end{figure}

One possible way to investigate whether there actually is any non-smooth behaviour
in the energy dependencies of particle yields, is the study of their local slope:
\begin{equation}
\frac{\Delta(N)}{\Delta(\sqrt{s_{_{\rbt{NN}}}})} = 
\frac{N_{\rb{i+1}} - N_{\rb{i}}}
{\sqrt{s_{_{\rbt{NN}}}}_{\rb{i+1}} - \sqrt{s_{_{\rbt{NN}}}}_{\rb{i}}}
\label{Eq:LocSl}
\end{equation}
This quantity is shown in \Fi{Fig:KPiDer} for charged pions and kaons, for which
the errors of the measurements are small enough to allow its calculation with
reasonable errors. The values of the local slopes have been derived from the
same data sets as shown in \Fi{Fig:Ratios}.
For the pions it decreases with increasing center-of-mass energy without exhibiting 
any significant structure. For kaons, on the other hand, there are indications
for sudden changes in the energy dependence around $\sqrts = 5 - 7$~GeV. The
local slope is clearly higher here compared to higher energies for the K$^{+}$. 
For the K$^{-}$ it seems to increase until $\sqrts \approx 7$~GeV and then does
not change any more and is also compatible to the local slopes for K$^{+}$ 
from there on.

\begin{figure}[t]
\begin{center}
\begin{minipage}[b]{70mm}
\begin{center}
\includegraphics[height=65mm]{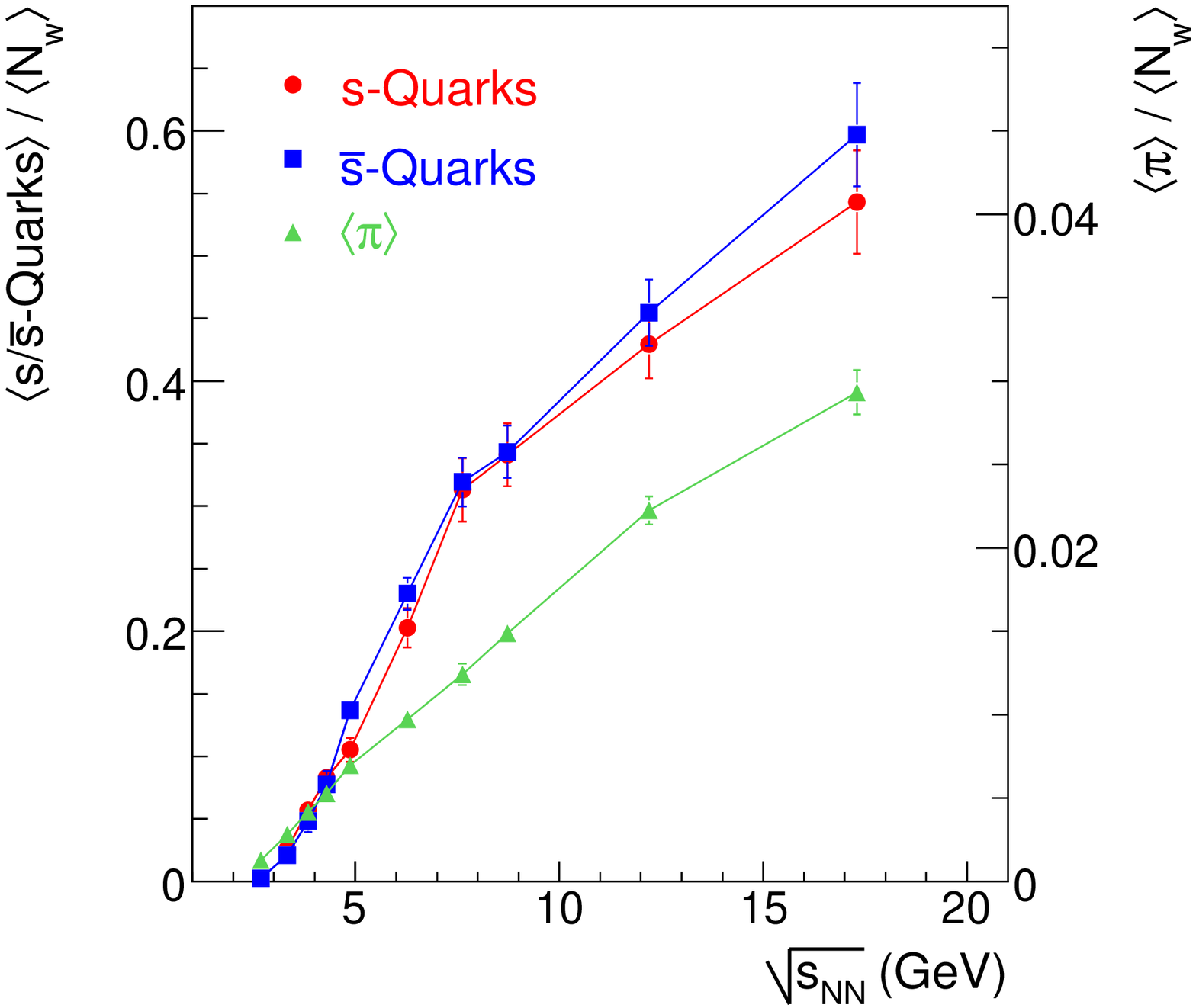}
\end{center}
\end{minipage}
\begin{minipage}[b]{70mm}
\begin{center}
\includegraphics[height=65mm]{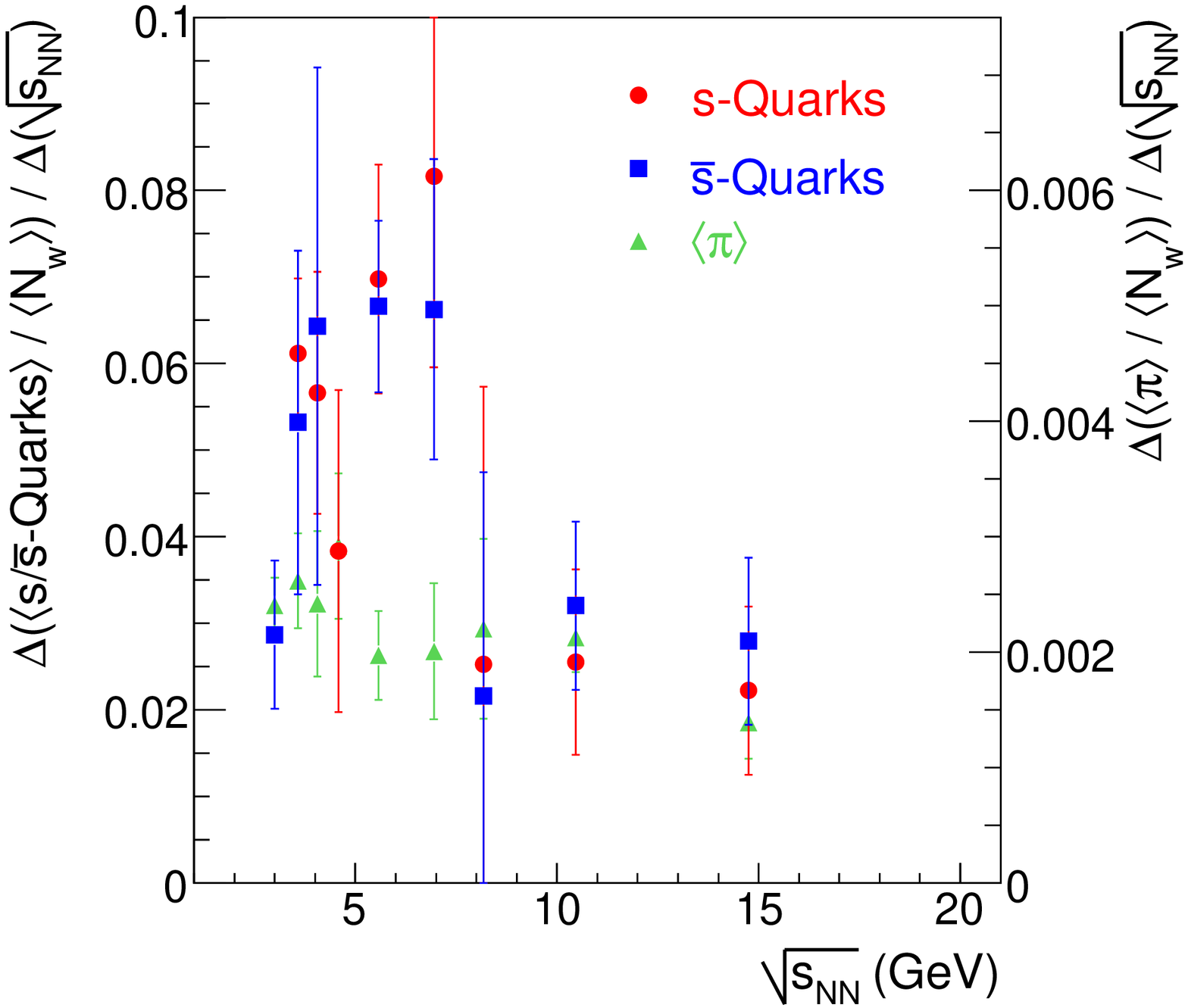}
\end{center}
\end{minipage}
\end{center}
\caption{Left: The total number of strange quarks and anti-quarks
as carried by kaons and hyperons together with the total number
of pions, both normalized to the number of wounded nucleons \nwound,
versus the collision energy for central Pb+Pb (Au+Au) reactions \cite{mysqm04}.
Right: The local slope of the energy dependencies of strange (anti-)quarks 
$\Delta \langle s/\bar{s}\textrm{-Quarks} / \nwound \rangle / \Delta \sqrts$
and pions 
$\Delta \langle \pi / \nwound \rangle / \Delta \sqrts$.}
\label{Fig:ssbar}
\end{figure}

From the measured total yields of the strange particles the
energy dependence of the number of produced strange quarks
and anti-quarks can be constructed \cite{mysqm04}.
The strange quark carriers which were taken into account are \kmin, \kzero, \lam\ 
(including \sigzero), \xis, \ommin, and \sigpm. For the strange anti-quark 
these are \kplus, \kzerob, \lab\ (including \sigzerob), \xisb, \omplus, and
\sigpmb \footnote{The \kzero\ contribution was calculated using isospin
symmetry ($\langle \kplus \rangle \approx \langle \kzero \rangle$,
$\langle \kmin \rangle \approx \langle \kzerob \rangle$). If no 
measurement is available, the values for the $\Xi$ and $\Omega$ 
yields were taken from statistical model fits \cite{becatt}. 
The \sigpm\ contribution was estimated based on the empirical factor
$(\langle \sigpm \rangle + \langle \lam \rangle)/\langle \lam \rangle = 1.6$
\cite{wrobl}.
Note that the strange quarks from the \myphi\ and $\eta$ are not included.}
The left panel of \Fi{Fig:ssbar} shows the energy dependence of the normalized 
number of strange (anti-)quarks in comparison to the total number of pions
($\langle\pi\rangle = 1.5 \left( \langle\pi^{+}\rangle + \langle\pi^{+}\rangle\right)$).
While $\langle\pi\rangle / \nwound$ is increasing with \sqrts\ rather smoothly, 
there is an indication for a change in the energy dependence of the total
strangeness production around $\sqrts \approx 7$~GeV. The significance of this
change can be visualized again by calculating the local slope, as defined in
\ref{Eq:LocSl}. While there is no change visible in the local slope for pions
(see right panel of \Fi{Fig:ssbar}), it is significantly higher for 
(anti-)strangeness below $\sqrts \approx 7$~GeV than above that energy, similar
to the behaviour already present for K$^{+}$.

\section{Transverse Mass Spectra}

\begin{figure}[htb]
\begin{center}
\includegraphics[width=130mm]{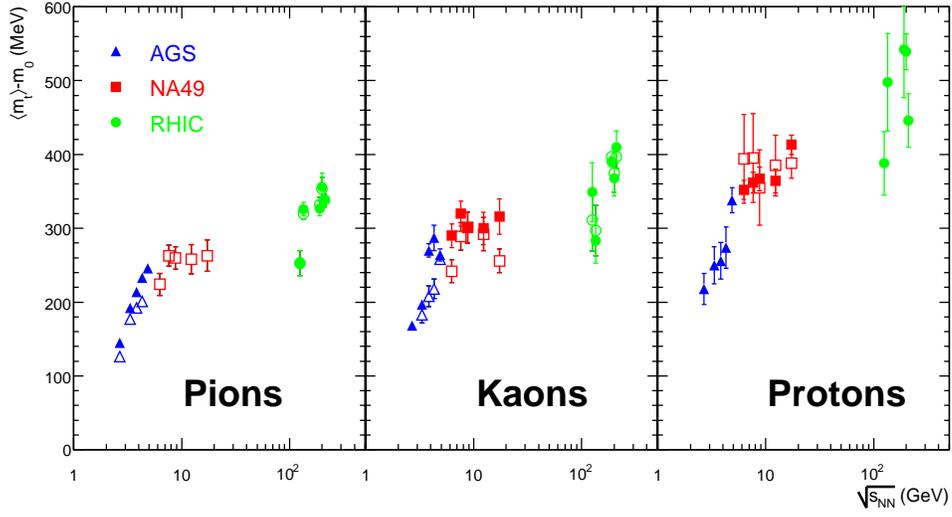}
\end{center}
\caption[]{The energy dependence of \meanmtm\ 
for pions, kaons, and protons at mid-rapidity for 5 (10\%) most central
Pb+Pb/Au+Au reactions. 
Open symbols represent negatively charged particles \cite{mysqm04}.}
\label{Fig:meanMt}
\end{figure}

\begin{figure}[htb]
\begin{center}
\includegraphics[width=130mm]{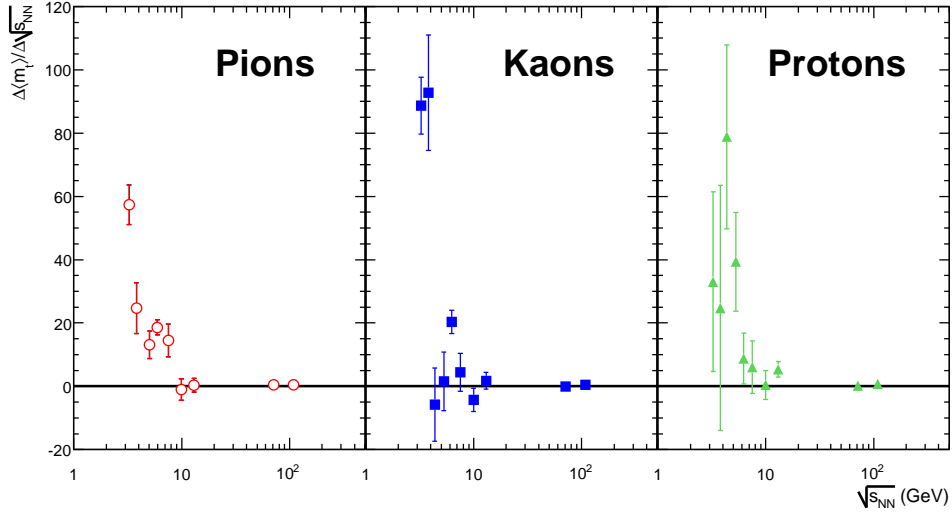}
\end{center}
\caption[]{The local slope of the energy dependence of the mean
transverse momentum
$\Delta \langle m_{\rb{t}} \rangle / \Delta \sqrts$ as a function 
of \sqrts\ for \pimin, \kplus, and protons.}
\label{Fig:meanMtder}
\end{figure}

The increase with energy of the inverse slope parameter $T$ of the kaon
\mt-spectra, as derived from an exponential fit, exhibits a sharp
change to a plateau around 30\agev\ \cite{marekt,mohanty}. Since the kaon \mt-spectra 
-- in contrast to the ones of the lighter pions or the heavier protons 
-- have to a good approximation an exponential shape, the inverse slope 
parameter provides in this case a good characterization of the spectra.
For other particle species, however, the local slope of the spectra depends
on \mt. Instead, the first moment of the \mt-spectra can be used
to study their energy dependence.
The dependence of \meanmtm\ on the center of mass energy \sqrts\ 
is summarized in \Fi{Fig:meanMt}. 
The change of the energy dependence around a beam energy of 
20 -- 30\agev\ is clearly visible for pions and kaons.
While \meanmtm\ rises steeply in 
the AGS energy range, the rise is much weaker from the low SPS energies 
on. To a lesser extent this change is also seen for protons.

Again the structures in the energy dependence can be investigated more
closely by calculation the local slope of the energy dependence of 
\meanmtm. Figure~\ref{Fig:meanMtder} shows 
$\Delta \langle m_{\rb{t}} \rangle / \Delta \sqrts$ as a function of
\sqrts. Generally, the local slope is higher at lower center-of-mass
energies and decreases ($\sqrts \leq 6 - 8$~GeV). It does not change 
any more when going to higher energies for all three particle species. 

\section{Elliptic Flow}

\begin{figure}[t]
\begin{center}
\begin{minipage}[b]{70mm}
\begin{center}
\includegraphics[width=70mm]{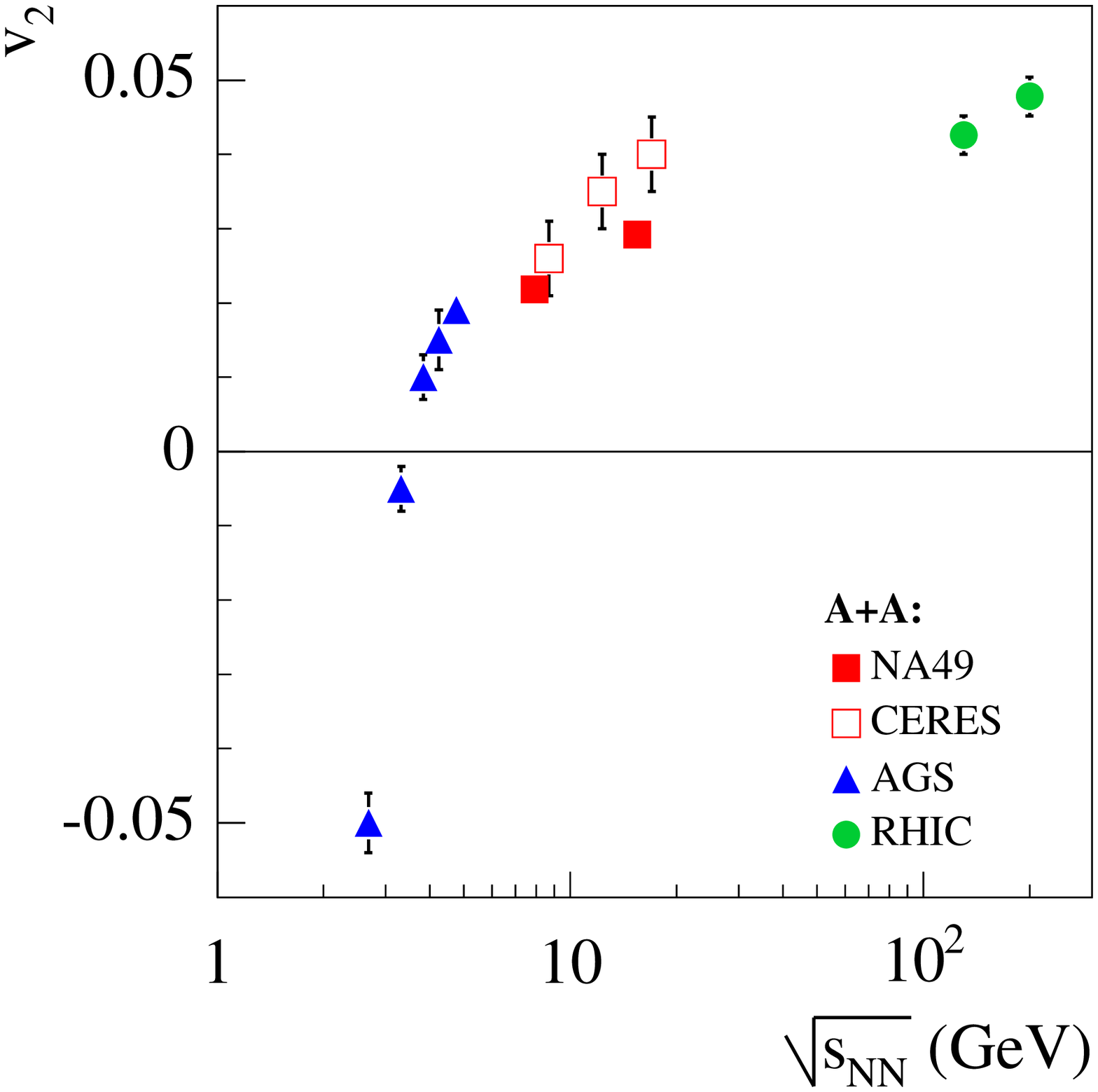}
\end{center}
\end{minipage}
\begin{minipage}[b]{70mm}
\begin{center}
\includegraphics[height=76mm]{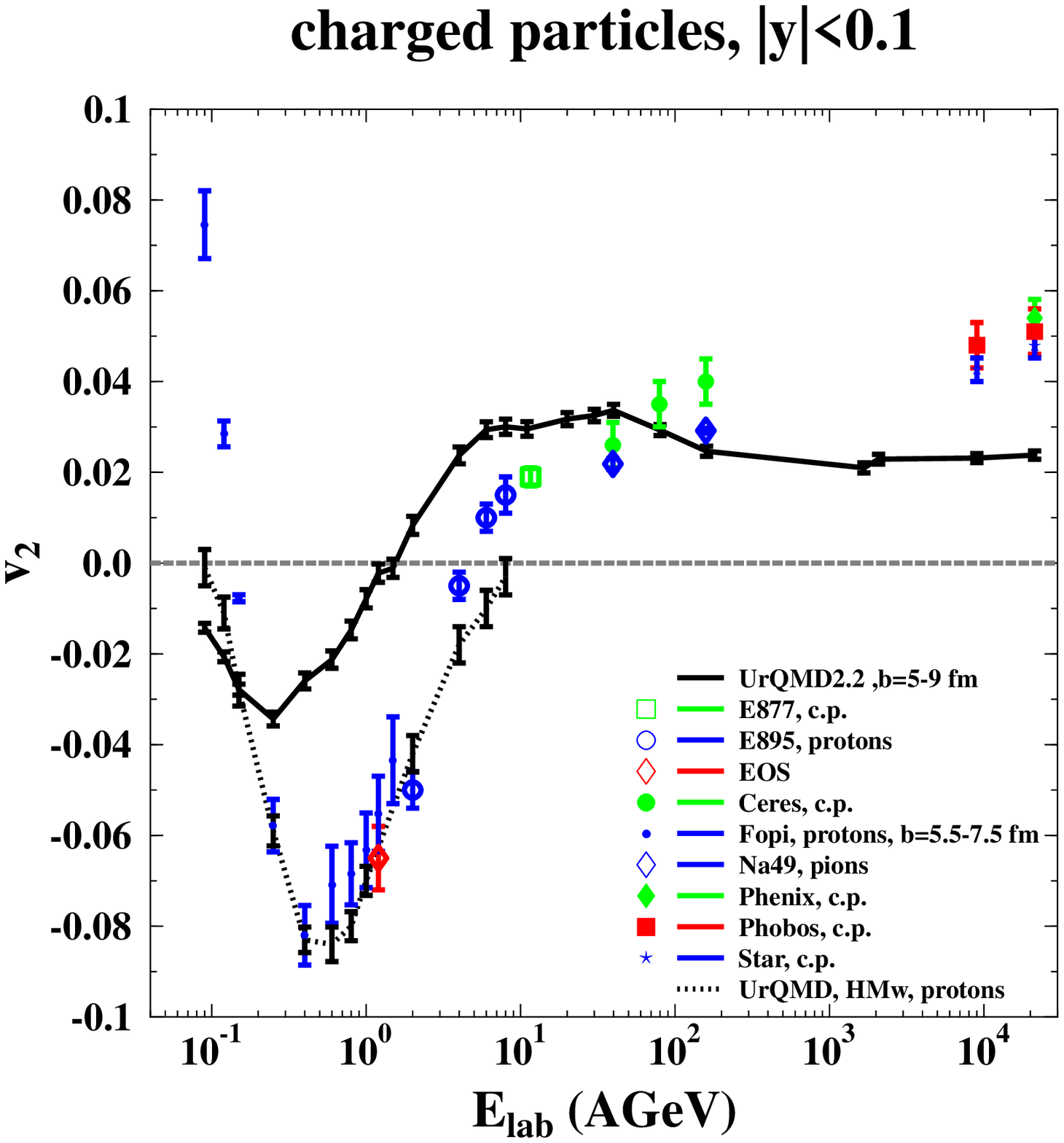}
\end{center}
\end{minipage}
\end{center}
\caption{Left: The energy dependence of \mypt-integrated values of 
\vtwo\ for pions at mid-rapidity \cite{agsfl1,agsfl2,na45fl,na49fl,rhicfl1,rhicfl2}.
Right: The energy excitation function of elliptic flow 
of charged particles in Au+Au/Pb+Pb collisions in mid-central 
collisions ($b = 5 - 9$~fm) with $|y|<0.1$ (full line) calculated with
the UrQMD model \cite{hannah}. 
This curve is compared to data from 
different experiments for mid-central collisions.}
\label{Fig:V2edep}
\end{figure}

An initial azimuthal anisotropy in space, as it is generated by 
the overlap region of two nuclei in non-central collisions, will 
be translated into a momentum anisotropy due to the different 
pressure gradients in the direction of the event plane and perpendicular
to it. The measurement of the second harmonic of 
this momentum anisotropy, \vtwo, provides thus a measure of the
initial pressure of the system. The energy dependence of \vtwo\
is summarized in the left panel of \Fi{Fig:V2edep}. 
One observes a rapid rise from negative \vtwo\ values at lower AGS 
energies to positive values, which changes its slope between 
top AGS and SPS energies. The further increase towards RHIC 
energies is then less pronounced. This measured energy dependence 
is compared to an UrQMD calculation in the right panel of 
\Fi{Fig:V2edep}, which is taken from \cite{hannah}. While the
energy dependence can roughly be described by this hadronic model 
at AGS energies and below, if a nuclear potential is included (dotted line), the
model results in too small values for \vtwo\ from SPS energies
on (solid line). This might indicate that for the higher energies
additional pressure from a partonic phase is needed in order
to decribe the data. According to this comparison the partonic 
contribution would set in at SPS energies. An exact determination
of the position of this onset would require more accurate data, 
since currently there is still a slight disagreement between the SPS data visible.

\section{Bose-Einstein Correlations}

\begin{figure}[htb]
\begin{center}
\includegraphics[width=162mm]{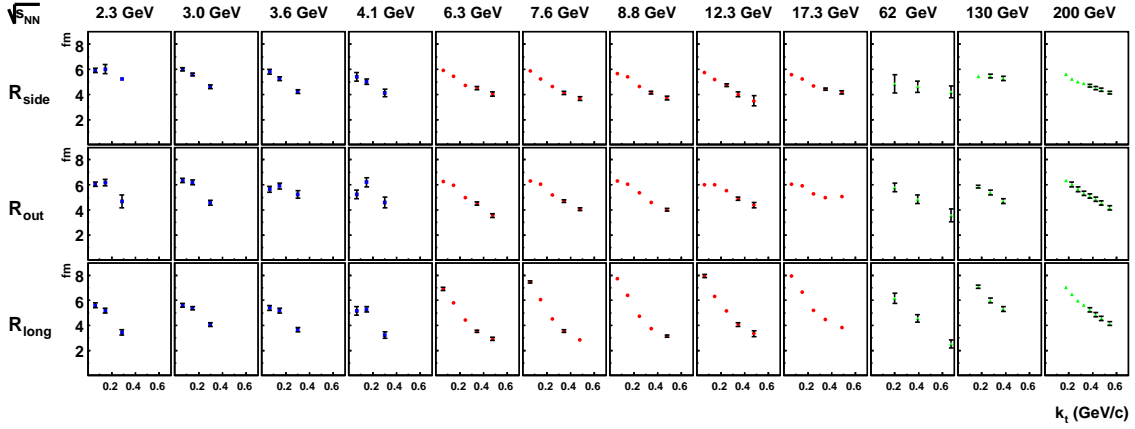}
\end{center}
\caption[]{HBT radius parameters as a function of $k_{\rb{t}}$ measured 
by E895 \cite{hbtags}, NA49 \cite{stefan}, PHOBOS \cite{hbtphbs}, 
and STAR \cite{hbtstar1,hbtstar2}.}
\label{Fig:HbtAll}
\end{figure}

\begin{figure}[htb]
\begin{center}
\includegraphics[width=110mm]{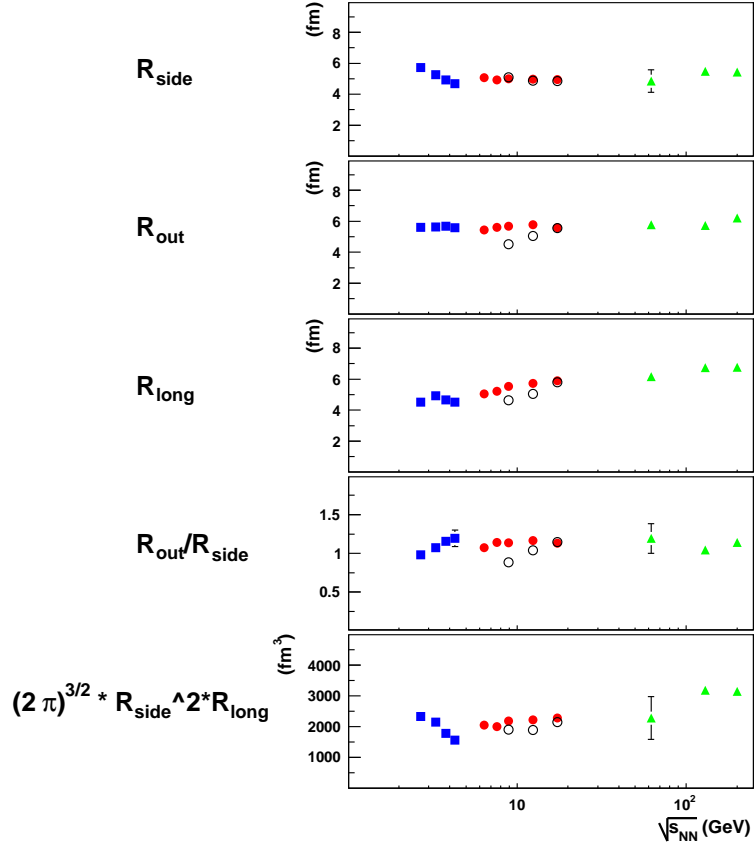}
\end{center}
\caption[]{The dependence of the various radius parameters on the center-of-mass
energy. The radii have been interpolated to $k_{\rb{t}} =$~0.2~\gevc\ \cite{stefan}.}
\label{Fig:HbtKt}
\end{figure}

Figure~\ref{Fig:HbtAll} shows a compilation of radius parameters 
measured in central nucleus nucleus collisions from AGS energies 
up to RHIC energies. One might expect that HBT radii should 
reflect also any change in the nature of the observed fireball,
for instance due to a change of the lifetime of the system.
However, even though other hadronic observable reveal some distinct
features in their energy dependence, as discussed in the previous
sections, structures in the energy dependence of HBT results are
remarkably absent. This is exemplified in \Fi{Fig:HbtKt} where the
\sqrts-dependence of different radius parameters at a fixed 
$k_{\rb{t}}$ is shown. While there are indications for a slight
variation at AGS energies, there is essentially no energy dependence
any more at higher energies with the only exception of $R_{\rb{long}}$,
which slowly rises. However, there are still some unresolved discrepancies
between NA49 \cite{stefan} and NA45 \cite{na45hbt}, where the latter 
measurement would suggest an
energy dependence of $R_{\rb{out}}$ at the SPS.

\section{Summary}

The study of the energy dependence of various hadronic observables
reveals several significant structures located in the energy region 
$\sqrts \approx 6 - 8$~GeV. There is an indication for a sudden
change in the strangeness production and the \meanmtm\ of pions, kaons, and
protons in this energy region, which can be identified by changes in
the local slopes. Similar structures might also be present in the
excitation function of elliptic flow, although here more measurements
with higher precision would be needed. On the other hand, there is only a
remarkably weak energy dependence of HBT-radii, with no clear signs 
of any structures.
The change in the strangeness production translates itself into a 
sharp maximum in the \kppip\ ratio. Generally, the ratios of strange
particles to pions are qualitativly described by a statistical hadron
gas model, assuming full chemical equilibrium. However, a detailed 
comparison to the measurements shows that there are significant differences
for $\sqrts > 7 - 8$~GeV which are about 3 - 5 times larger than the total error.
(kaons, $\phi$, and $\Xi$). However, in order to establish whether
this indicates the onset of deconfinement, more measurement with
high precision would be highly desirable.


\begin{thebibliography}{99}

\bibitem{michi}    M.~Mitrovski et al. (for the NA49 collaboration),
                   arXiv:nucl-ex/0606004.

\bibitem{marekqm}  M.~Ga\'zdzicki (for the NA49 collaboration), 
                   J. Phys. G {\bf 30}, s701 (2004).

\bibitem{smes}     M.~Ga\'{z}dzicki and M.I.~Gorenstein,
                   Acta Phys. Polon. B {\bf 30}, 2705 (1999).

\bibitem{cleymans} J.~Cleymans, H.~Oeschler, K.~Redlich, and S.~Wheaton,
                   Eur. Phys. J. A {\bf 29}, 119 (2006).

\bibitem{urqmd}    M.~Bleicher et al.,
                   J. Phys. G {\bf 25}, 1859 (1999).

\bibitem{hsd}      E.L.~Bratkovskaya et al.,
                   Phys. Rev. C {\bf 69}, 054907 (2004).

\bibitem{pbm}      P.~Braun-Munzinger, J.~Cleymans, H.~Oeschler, and K.~Redlich,
                   Nucl. Phys. A {\bf 697}, 902 (2002).

\bibitem{becatt}   F.~Becattini, M.~Ga\'zdzicki, A.~Ker\"anen, J.~Manninen, and R.~Stock,
                   Phys. Rev. C {\bf 69}, 024905 (2004).

\bibitem{pbmjs}    P.~Braun-Munzinger, J.~Stachel, C.~Wetterich,
                   Phys. Lett. B {\bf 596}, 61 (2004).

\bibitem{wrobl}    A.K.~Wr\'{o}blewski,
                   Acta Phys. Polon. B {\bf 16}, 379 (1985).

%
%
%
%
%
%
%
%
%
%
%

\bibitem{mysqm04}  C.~Blume (for the NA49 collaboration),
                   J. Phys. G {\bf 31}, s685 (2005).

\bibitem{marekt}   M.I.~Gorenstein, M.~Ga\'zdzicki, and K.A.Bugaev,
                   Phys. Lett. B {\bf 567}, 175 (2003).

\bibitem{mohanty}  B.~Mohanty, J.~Alam, S.~Sarkar, T.K.~Nayak, and B.K.~Nandi,
                   Phys. Rev. C {\bf 68}, 021901 (2003). 

\bibitem{agsfl1}   C.~Pinkenburg et al. (E895 collaboration),
                   Phys. Rev. Lett. {\bf 83}, 1295 (1999).

\bibitem{agsfl2}   J.~Barrette et al. (E877 collaboration),
                   Nucl. Phys. A {\bf 661}, 329c (1999).

\bibitem{na45fl}   J.~Slivova et al. (NA45/CERES collaboration),
                   Nucl. Phys. A {\bf 715}, 615c (2003).

\bibitem{na49fl}   C.~Alt et al. (NA49 collaboration),
                   Phys. Rev. C {\bf 68}, 034903 (2003).

\bibitem{rhicfl1}  C.~Adler et al. (STAR collaboration),
                   Phys. Rev. Lett. {\bf 87}, 182301 (2001).

\bibitem{rhicfl2}  J.~Adams et al. (STAR collaboration),
                   Phys. Rev. C {\bf 72}, 014904 (2005).

\bibitem{hannah}   H.~Petersen, Q.~Li, X.~Zhu, and M.~Bleicher,
                   arXiv:hep-ph/0608189.

\bibitem{hbtags}   M.~Lisa et al. (E895 collaboration),
                   Phys. Rev. Lett. {\bf 84}, 2798 (2000).

\bibitem{hbtphbs}  B.B.~Back et al. (PHOBOS collaboration),
                   Phys. Rev. C {\bf 73}, 031901(R) (2006).

\bibitem{hbtstar1} C.~Adler et al. (STAR collaboration),
                   Phys. Rev. Lett. {\bf 87}, 082301 (2001).

\bibitem{hbtstar2} J.~Adams et al. (STAR collaboration),
                   Phys. Rev. C {\bf 71}, 044906 (2005).

\bibitem{stefan}   S.~Kniege et al. (NA49 collaboration),
                   J. Phys. G {\bf 30}, s1073 (2004).

\bibitem{na45hbt}  D.~Adamova et al. (NA45/CERES collaboration),
                   Nucl. Phys. A {\bf 714}, 124 (2003).

\end{thebibliography}
\end{document}